\documentclass[prl,showpacs,twocolumn]{revtex4}
\usepackage{graphicx,psfrag,amssymb,amsmath}

\begin{document}
\title{Magnetothermopower and Nernst effect in unconventional charge density waves}

\author{Bal\'azs D\'ora$^1$, Kazumi Maki$^2$, Andr\'as V\'anyolos$^3$ and Attila Virosztek$^{3,4}$}

\address{$^1$ The Abdus Salam ICTP, Strada Costiera 11, I-34014, Trieste, Italy} 
\address{$^2$ Department of Physics and Astronomy, University of Southern
California, Los Angeles CA 90089-0484, USA}  
\address{$^3$ Department of Physics, Budapest University of Technology and
Economics, H-1521 Budapest, Hungary}
\address{$^4$ Research Institute for Solid State Physics and Optics, P.O. Box 49, H-1525 Budapest, Hungary}

\date{\today}

\begin{abstract}

Recently we have shown that the striking angular dependent
magnetoresistance in the low temperature phase (LTP) of
$\alpha$-(BEDT-TTF)$_2$KHg(SCN)$_4$ is consistently described in terms of
unconventional charge density wave (UCDW).
Here we investigate theoretically the thermoelectric power and the Nernst
effect in UDW. The present results account consistently for the recent
data of magnetothermopower in $\alpha$-(BEDT-TTF)$_2$KHg(SCN)$_4$ obtained
by Choi et al. (Phys. Rev. B, \textbf{65}, 205119 (2002)).
This confirms further our identification of LTP in this salt as UCDW. We
propose also that the Nernst effect provides a clear signature of UDW.
\end{abstract}


\maketitle


Recently many possible candidates for unconventional charge density
wave (UCDW) and unconventional spin density wave (USDW) have been
proposed, though in most cases definitive confirmation is still lacking.
These are the antiferromagnetic phase
of URu$_2$Si$_2$\cite{IO,roma}, the pseudogap
phase in high $T_c$ cuprates\cite{benfatto,Sudip,nayak,krakko}, the CDW in 
NbSe$_2$\cite{castroneto,bel} and the low temperature
phase (LTP) in $\alpha$-(BEDT-TTF)$_2$MHg(SCN)$_4$ with M=K, Rb and 
Tl\cite{kuszobter,rapid,tesla,epladmr,admrprl}.
In the last system not only the qualitative features of LTP, like the
absence of a clear charge order, but also both the temperature and magnetic
field dependence of the threshold electric field\cite{kuszobter,rapid,tesla} and the striking 
angular
dependent magnetoresistance (ADMR) \cite{epladmr,admrprl} are fully consistent with UCDW. In these
studies the quantization of the quasiparticle spectrum in the presence of
magnetic field as considered by Nersesyan et al. \cite{Ners1,Ners2} plays the crucial role.

The object of the present paper is to extend our earlier study to the
thermoelectric power and Nernst effect in UDW (i.e. UCDW and USDW) in the
presence
of magnetic field. When the Zeeman splitting or the Pauli term due to
magnetic field is negligible compared to the orbital effect, there will be
no distinction between UCDW and USDW, which we will assume in the
followings. First we discuss briefly how the effect of magnetic
field is incorporated following Refs. \cite{Ners1,Ners2}.
Then we construct the expressions for thermopower and Nernst
effect in UDW. These are compared with a recent data by Choi et al.\cite{choi} on
$\alpha$-(BEDT-TTF)$_2$KHg(SCN)$_4$. Indeed we can describe the
experimental data very consistently.


In the absence of magnetic field the quasiparticle energy in UCDW is given 
by\cite{makiegyedul}
\begin{equation}
(E+\varepsilon({\bf k}))\Psi=(-iv_a\partial_x\rho_3+\Delta\cos(ck_z)\rho_1)\Psi,
\label{alap}
\end{equation}
where $\rho_i$'s are the Pauli matrices acting on spinor space of the left 
and right moving electrons on the
quasi-one dimensional Fermi surface and the imperfect nesting term $\varepsilon(\bf k)$ is given 
by\cite{admrprl}
\begin{equation}
\varepsilon({\bf k})=\sum_{n=-\infty}^\infty \varepsilon_n\cos(2{\bf b}_n^\prime{\bf k}),
\end{equation}
where ${\bf b}^\prime_n=b^\prime[\hat{\bf
r}_b+\tan(\theta_n)(\hat{\bf r}_a\cos\phi_0+\hat{\bf r}_c\sin\phi_0)]$, $\varepsilon_n=\varepsilon_0 2^{-|n|}$,
$\tan(\theta_n)=\tan(\theta_0)+n
d_0$, $\tan(\theta_0)\simeq 0.5$, $d_0\simeq 1.25$,
$\phi_0\simeq 27^\circ$\cite{kovalev,fermi,hanasaki}, and $\phi$ is the angle the projected
magnetic field on the
$a-c$ plane
makes from the $c$-axis.
This generalized
imperfect nesting term arises from an effective tight binding approximation, where hopping takes place between sites in 
the
$\hat{\bf r}_b$ direction and along nearest neighbour chains oriented in the $\hat{\bf r}_a\cos\phi_0+\hat{\bf r}_c\sin\phi_0$
direction.
Eq. (\ref{alap}) is readily solved as
\begin{equation}
E=\pm\sqrt{(v_ak_x)^2+\Delta^2\cos^2(ck_z)}-\varepsilon(\bf k),\label{elso}  
\end{equation}
where $k_x$, $k_z$ are wavevector components parallel to the $a$ and $c$ axis in
$\alpha$-(BEDT-TTF)$_2$KHg(SCN)$_4$ salt.
We note here that imperfect nesting breaks the particle-hole symmetry in general.
In the presence of magnetic field Eq. (\ref{alap}) is transformed as
\begin{equation}
E\Psi=(-iv_a\partial_x\rho_3+\Delta ceBx\cos(\theta)\rho_1)\Psi,
\label{ujalap}
\end{equation}  
where for the moment we ignored the imperfect nesting term. $\theta$ is the angle the magnetic
field makes from the $b^*$ axis. We define $b^*$ as the direction perpendicular to the $a-c$ 
plane. 
Eq. (\ref{ujalap}) is readily solved as\cite{Ners1,Ners2}
\begin{equation}
E^2=2nv_a\Delta c e |B\cos\theta|,
\end{equation}
where $n=0$, $1$, $2$\dots.
We note that Eq. (\ref{ujalap}) is the same as the Dirac equation in a constant magnetic field and 
has been studied since 1936\cite{heisenberg}. The Landau wavefunctions are given by
\begin{gather}
\Psi_0=\left( \begin{array}{c}
i\\
1
\end{array}\right)\phi_0, \\
\Psi_{n\neq 0}=\frac{1}{\sqrt{2}}\left[\left(\begin{array}{c}
1\\
i
\end{array}\right)\phi_{n-1}\pm
\left(\begin{array}{c} 
i\\
1
\end{array}\right)\phi_{n}\right],
\label{hullamfgv}
\end{gather}
where $\phi_n$ is the $n$-th wavefunction of a linear harmonic oscillator with parameters 
"mass" $m=1/2v_a^2$ and "frequency" $\omega=2 v_a \Delta ceB\cos(\theta)$. From Eq.
(\ref{hullamfgv}) it is obvious, that the $n\neq 0$ levels are 
twofold degenerate, since $\Psi_{n\neq 0}$ is
composed of
the
$n-1$-th and $n$-th wavefunction of the harmonic oscillator. 
Now making use of the Landau wave functions we evaluate the
contribution from
the imperfect nesting
term as perturbation. Then we get for the Landau levels:
\begin{gather}
E_{0,1}=-E_0^{(1)},\\
E_{1,1}=\pm E_1-E_1^{(1)},\\
E_{1,2}=\pm E_1-E_1^{(2)},
\end{gather}
and
\begin{gather}
E_n=\sqrt{2nv_a\Delta ceB |\cos(\theta)|},\\
E_0^{(1)}=E_1^{(1)}=\sum_m\varepsilon_m\exp(-y_m),\\
E_1^{(2)}=\sum_m\varepsilon_m(1-2y_m)\exp(-y_m),
\end{gather}
where $y_m=v_a {b^\prime}^2e |B\cos(\theta)|(\tan(\theta)\cos(\phi-\phi_o)
-\tan(\theta_m))^2/\Delta c$.
We note that the imperfect nesting terms splits the $n=1$ Landau level into 2 nondegenerate levels.
Also as is the case in the absence of magnetic field, the imperfect nesting term breaks the 
particle-hole symmetry. This particle-hole symmetry breaking is crucial in the thermoelectric
power.
Then keeping just the $n=0$ and $n=1$ Landau levels, the ADMR is constructed as
\begin{gather}
R(B,\theta,\phi)^{-1}=\sigma_1\left(\dfrac{\exp(-x_1)+\cosh(\zeta_0)}{\cosh(x_1)+\cosh(\zeta_0)}+\right.\nonumber \\
\left.+\dfrac{\exp(-x_1)+\cosh(\zeta_1)}{\cosh(x_1)+\cosh(\zeta_1)}\right)+\sigma_2,
\label{fit}
\end{gather}
where $x_1=\beta E_1$, $\zeta_0=\beta E_1^{(1)}$, $\zeta_1=\beta E_1^{(2)}$ and $1/\beta=k_B T$.
Here $\sigma_1$ is the conductivity of the $n=1$ Landau level and the contribution from the 
$n=0$ Landau level, $\sigma_0$ (which was found to be
constant within the present approximation) is considered together with the conductivity of 
the quasi-two dimensional Fermi surface in $\sigma_2$, which is also assumed to be independent of
$B$.
Also from our construction of ADMR, Eq. (\ref{fit}) should work better for smaller $T$ 
and larger $B$. As was shown in Ref. \cite{epladmr,admrprl}, Eq. (\ref{fit}) gives an excellent 
description of ADMR
found in $\alpha$-(BEDT-TTF)$_2$KHg(SCN)$_4$.

The diagonal component of the magnetic field dependent thermoelectric power (TEP) is also of
particular interest. It is obtained similarly as Eq. (\ref{fit}),
and reads:
\begin{gather}
S(B,\theta,\phi)=-\frac{R(B,\theta,\phi)k_B}{e}\left[\sigma_0 \zeta_0+\right.\nonumber\\
+\left.\sigma_1\left(\zeta_0
\frac{\exp(-x_1)+\cosh(\zeta_0)}{\cosh(x_1)+\cosh(\zeta_0)}+
\zeta_1 \frac{\exp(-x_1)+\cosh(\zeta_1)}{\cosh(x_1)+\cosh(\zeta_1)}+\right.\right.\nonumber \\
+\left.\left.x_1\left(
\frac{\sinh(\zeta_0)}{\cosh(x_1)+\cosh(\zeta_0)}+ \frac{\sinh(\zeta_1)}{\cosh(x_1)+\cosh(\zeta_1)}\right)\right)\right] 
\label{tep}
\end{gather}
We note here that $S(B,\theta,\phi)$ vanishes in the absence of imperfect nesting.
Before comparing Eq. (\ref{tep}) with experimental data, we shall consider the Nernst effect.


The Nernst effect is the off diagonal component of the thermoelectric power in the presence of
magnetic field. Also its formulation is different from above. We have seen already that
quasiparticle in UDW orbits around the magnetic field. Then when an electric field $\bf E$ is
applied with a perpendicular component to the magnetic field $\bf B$, the quasiparticle orbit
drifts with ${\bf v}_D=({\bf E\times B})/B^2$. Then the heat current parallel to ${\bf v}_D$ is
given by ${\bf J}_h=TS{\bf v}_D$, where $S$ is the entropy associated with the circling
quasiparticles:
\begin{equation}
S=eB\sum_n\left[\ln(1+\exp(-\beta E_n))+\beta E_n(1+\exp(\beta 
E_n))^{-1}\right],
\label{entr}
\end{equation} 
the sum over $E_n$ has to be taken over all the Landau levels, and the magnetic field is assumed to be perpendicular to the $a-c$ 
plane ($\theta=0^\circ$). Then for small $T$ and large $B$,
Eq. (\ref{entr}) is well approximated by taking the $n=0$ and $n=1$ Landau levels. 
Moreover, when the zeroth order contribution from the energy spectrum (i.e. the Landau levels without 
imperfect nesting) is finite, we can neglect higher order terms, namely the effect of imperfect 
nesting by setting $\zeta_0=\zeta_1=0$. With this simplification, the entropy reads as 
\begin{equation}
S=2eB\left[\ln(2)+2\ln\left(2\cosh\left(\frac{x_1}{2}\right)\right)
-x_1\tanh\left(\frac{x_1}{2}\right)\right].
\label{ner}
\end{equation}
So the Nernst coefficient in this configuration can be calculated, after considering the effect of the two dimensional parts 
of the Fermi surface:
\begin{gather}
S_{xy}=\alpha_{xy}=-\frac{S}{B\sigma}=\frac{1}{\sigma}\left[\dfrac{L_{\textmd{2D}}}{1+\gamma^2 
B^2}-\right.\nonumber\\
\left.-2e\left(\ln(2)+2\ln\left(2\cosh\left(\frac{x_1}{2}\right)\right)
-x_1\tanh\left(\frac{x_1}{2}\right)\right)\right],
\label{nern}
\end{gather}
where $\sigma=1/R=4\sigma_1/(\exp(x_1)+1)+\sigma_2$ from Eq. (\ref{fit}), $L_{\textmd{2D}}$ stems from the two dimensional cylinders 
of 
the Fermi 
surface, $\gamma=e \tau/m$, 
$\tau$ is the field-free relaxation time at the Fermi level\cite{ner1,ner2}, $m$ is the effective mass of the electron.

\begin{figure}[h!]
\psfrag{x}[t][b][1.2][0]{$B$(T)}
\psfrag{y}[b][t][1.2][0]{$S$($\mu$V/K)}
\psfrag{x1}[l][r][1][0]{$T=1.4$~K}
\psfrag{x2}[r][l][1][0]{$T=4.8$~K}
\psfrag{x3}[l][][1][0]{$T=5.8$~K}
\psfrag{x4}[t][b][1][0]{$T=6.9$~K}
\includegraphics[width=7cm,height=7cm]{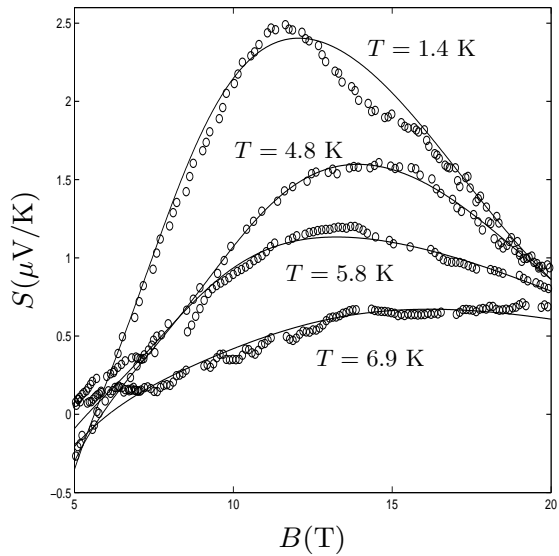}
\caption{The magnetothermopower for heat current along the $a$ direction is shown for $T=1.4$~K, 
$T=4.8$~K ,$T=5.8$~K and $T=6.9$~K 
from top to bottom,
the circles denote the experimental data from Ref. \cite{choi}, the solid line is our fit based 
on Eq. (\ref{tep}).}\label{choi4a}
\end{figure}

\begin{figure}[h!]
\psfrag{x}[t][b][1.2][0]{$B$(T)}
\psfrag{y}[b][t][1.2][0]{$S$($\mu$V/K)}
\psfrag{x5}[][][1][0]{$T=0.7$~K}
\includegraphics[width=7cm,height=7cm]{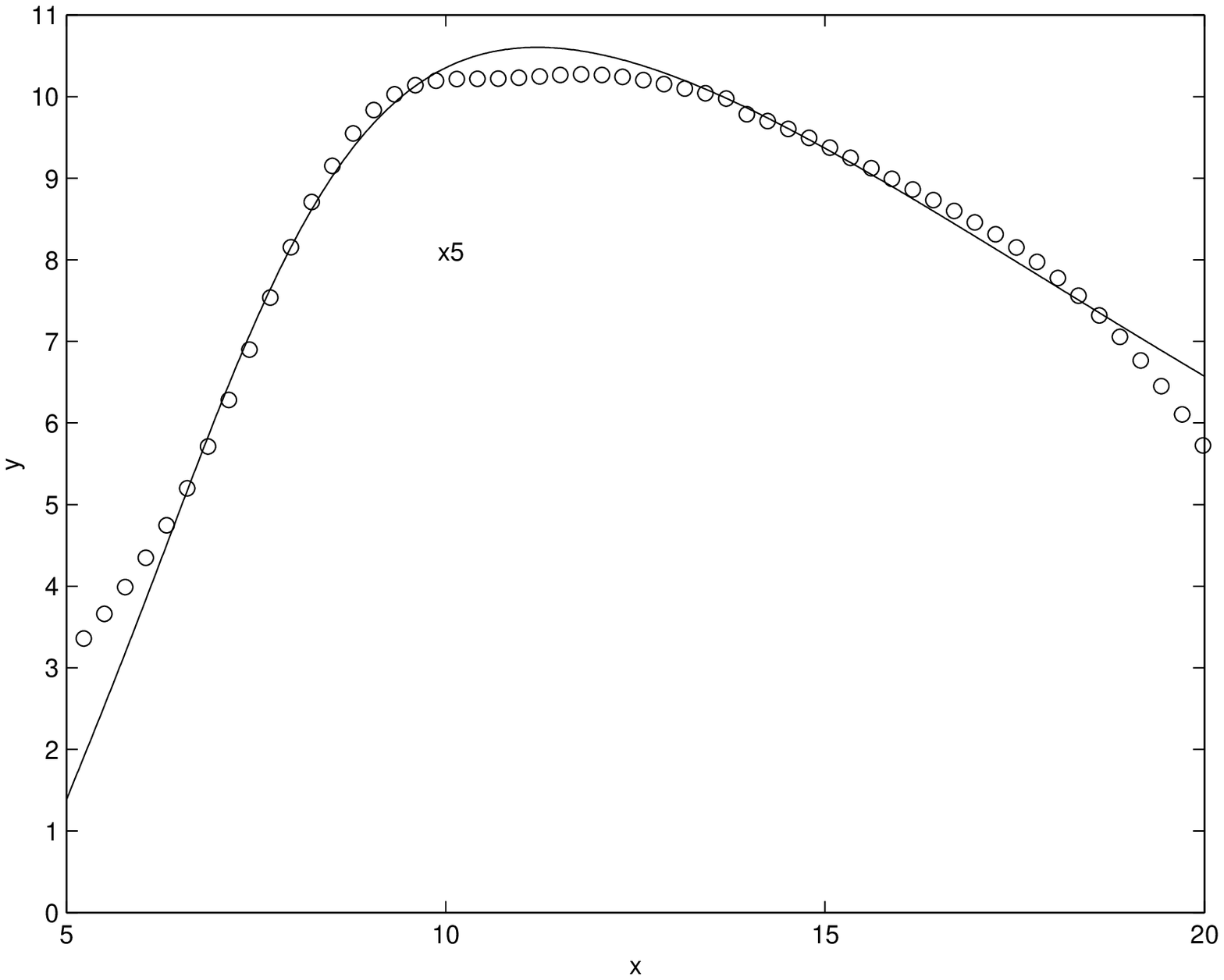}
\vspace*{6mm}

\psfrag{x}[t][b][1.2][0]{$B$(T)}
\psfrag{y}[b][t][1.2][0]{$S$($\mu$V/K)}
\psfrag{x6}[][][1][0]{$T=1.5$~K}
\includegraphics[width=7cm,height=7cm]{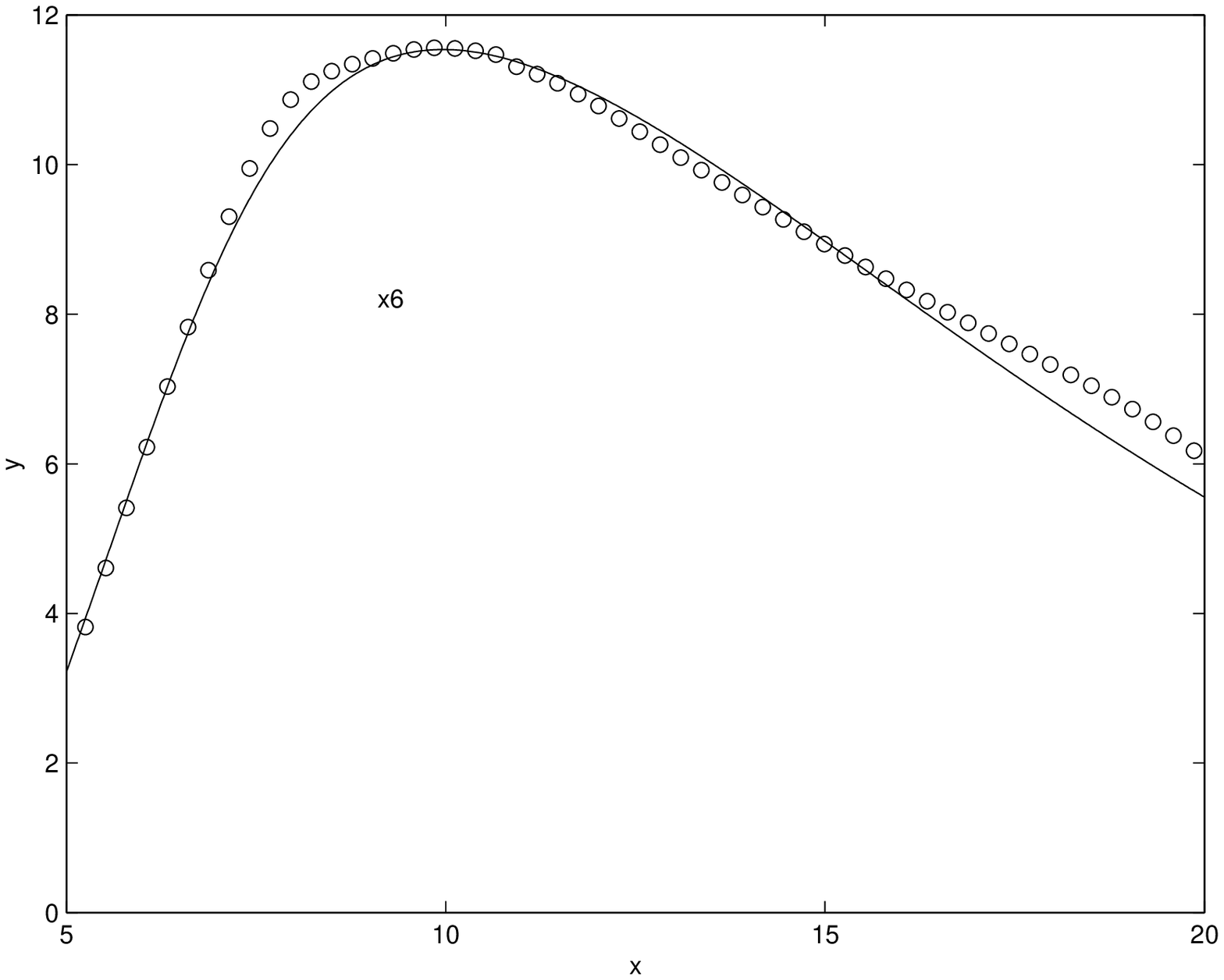}
\caption{The magnetothermopower for heat current along the $c$ direction is shown for $T=0.7$~K 
(upper panel) and  $T=1.5$~K (lower panel), the circles denote the experimental data from Ref. 
\cite{choi}, the solid line is our fit based on Eq. (\ref{tep}).}\label{choi4b}
\end{figure}

\begin{figure}[h!]
\centering
\psfrag{x}[t][b][1.2][0]{$B$(T)}
\psfrag{y}[b][t][1.2][0]{$S_{xy}$($\mu$V/K)}
\psfrag{x6}[l][][1][0]{$T=1.4$~K}
\psfrag{x7}[l][][1][0]{$T=4.8$~K}
\includegraphics[width=7cm,height=7cm]{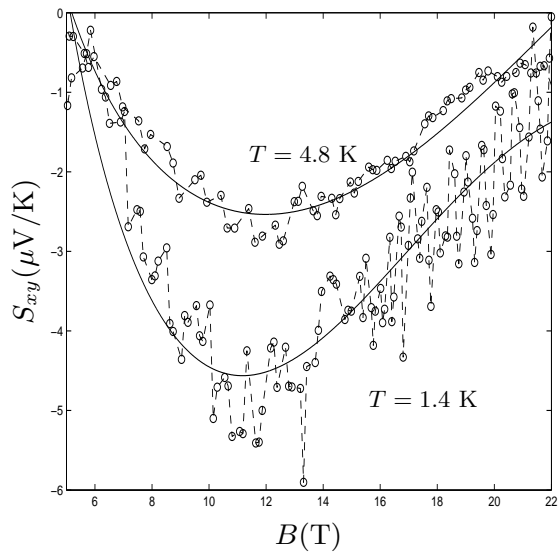}
\caption{The Nernst signal for heat current along the $a$ direction is shown 
for $T=1.4$~K and  $T=4.8$~K (from bottom to top), the dashed lines with circles denote the 
experimental data from Ref. \cite{choi}, the solid line is our fit based on Eq. 
(\ref{nern}).}\label{choi5a}
\end{figure}

\begin{figure}[h!]
\psfrag{x}[t][b][1.2][0]{$T$(K)}
\psfrag{y}[b][t][1.2][0]{$S$($\mu$V/K)}
\psfrag{x8}[][][1][0]{$B=12$~T}
\includegraphics[width=7cm,height=7cm]{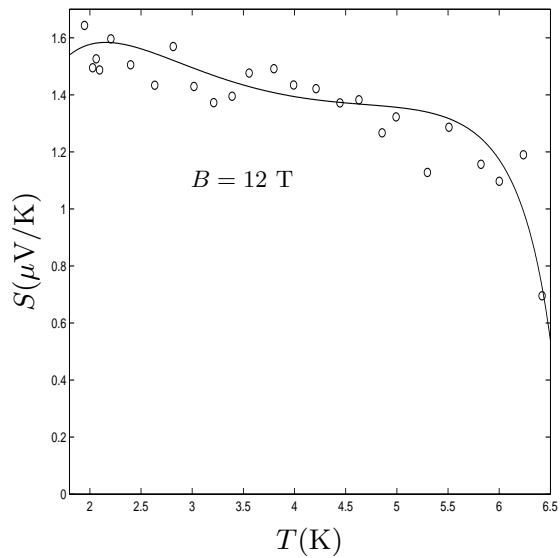}
\caption{The temperature dependence of the magnetothermopower for heat current along the $a$ 
direction is shown for $B=12$~T, the circles denote the experimental data from Ref.
\cite{choi}, the solid line is our fit based on Eq. (\ref{tep}).}\label{choi6a}
\end{figure}

In Figs. \ref{choi4a} and \ref{choi4b},  we compare Eq. 
(\ref{tep}) to the experimental data of the diagonal thermopower (Seebeck coefficient) in a magnetic field perpendicular to the 
conducting plane ($\theta=0^\circ$) for 
heat current applied in the $a$ and $c$ directions, respectively.
As is 
readily seen we can have excellent fittings. From these fittings we can deduce 
$\Delta\sim17$~K,
$v_a\sim 10^6$~cm/s, which are very consistent with earlier results. 
As to $\varepsilon_0$, its value is obtained as $9$~K for $T=1.4$~K, which is of the same order of magnitude as what we obtained 
previously in 
Ref. \cite{admrprl}, but for higher temperatures (above half $T_c$) we obtain $\varepsilon_0$ around $40$~K. This bigger value might 
stem  
from the neglect of the 
effect of higher Landau levels (which becomes more important as the temperature increases), and also of the magnetic field 
and temperature 
dependence of $\sigma_1$ and $\sigma_2$.
In Fig. \ref{choi5a}, 
we compare our theoretical results to the experimental data on Nernst effect. From this we can 
deduce the same parameters, and the new fitting parameter, $\tau=10^{-11}$~s, assuming $m$ to be twice the free electron mass.
This can be converted to temperatures as $\tau^{-1}\rightarrow 4$~K, which is reasonable concerning the presence of de Haas van Alphen 
oscillations at $B>10$~T at $T=1.4$~K.
Finally in Fig. \ref{choi6a}, the temperature dependence of the Seebeck coefficient is fitted 
at $B=12$~T. Here we assumed $\Delta(T)/\Delta(0)=\sqrt{1-(T/T_c)^3}$, which is very close to 
our weak coupling solution\cite{nagycikk}. The extracted fitting parameters are again in the 
same order of magnitude as earlier.

As we have shown the quasiparticle spectrum as obtained can describe the 
magnetothermopower as observed in LTP in  $\alpha$-(BEDT-TTF)$_2$KHg(SCN)$_4$. Together with 
the earlier results on ADMR, the present work further confirms our proposition that LTP in 
$\alpha$-(BEDT-TTF)$_2$KHg(SCN)$_4$ salt is UCDW.
Also the Nernst effect we obtained is rather large and independent of the imperfect nesting term. 
Therefore we may consider the Nernst effect as 
the hallmark of UDW. 
No corresponding
term exists in conventional DW.
The $B$ dependence of the Nernst effect is very similar to the one 
obtained in the vortex state of dirty type II s-wave 
superconductors\cite{caroli1,caroli2,caroli3,caroli4}. It has been claimed by 
Wang et al.\cite{ong}, that the large Nernst and Ettinghausen effect in the pseudogap phase is 
the 
signature of the presence of superconducting vortices. But the present results point clearly to 
the alternative possibility. Indeed the beautiful experimental data from underdoped LSCO 
appear to indicate that the pseudogap phase is UDW as proposed by many people\cite{benfatto,Sudip,nayak,krakko}. Of course the 
quantitative comparison between experiment and theory as done here for LTP in 
$\alpha$-(BEDT-TTF)$_2$KHg(SCN)$_4$ is highly desirable. We also believe that measurements of Nernst and/or 
Ettinghausen effect will prove to be decisive in other possible UDW candidate materials.

We would like to thank Eun Sang Choi for sending us the data of Ref. \cite{choi}. 
We have benefited from discussions with Mario Basleti\'c, Mark Kartsovnik, Bojana 
Korin-Hamzi\'c, Amir Hamzi\'c, Silvia Tomi\'c and Peter Thalmeier.
We thank Yuji Matsuda for drawing our attention to Ref. \cite{bel}. Two of us (K.M. and B.D.) 
acknowledge the hospitality and support of the Max Planck Institute for the
Physics of Complex Systems and Max Planck Institute for Chemical Physics of Solids, Dresden, 
where part of this work was done.
This work
was supported by the Hungarian National Research Fund under grant numbers
OTKA T032162 and NDF45172.

\bibliographystyle{apsrev}
\bibliography{mr}
\end{document}